\documentstyle [12pt] {article}
\makeatletter

\@addtoreset{equation}{section}
\ifcase \@ptsize
\or % mods for 11 pt
\or % mods for 12 pt
\fi
\advance\textheight by \topskip

\ifcase \@ptsize
\or % mods for 11 pt
\or % mods for 12 pt
\fi

\def\@citex[#1]#2{%
\if@filesw \immediate \write \@auxout {\string \citation {#2}}\fi
\@tempcntb\m@ne \let\@h@ld\relax \def\@citea{}%
\@cite{%
  \@for \@citeb:=#2\do {%
    \@ifundefined {b@\@citeb}%
      {\@h@ld\@citea\@tempcntb\m@ne{\bf ?}%
      \@warning {Citation `\@citeb ' on page \thepage \space undefined}}%
      {\@tempcnta\@tempcntb \advance\@tempcnta\@ne%
      \@tempcntb\number\csname b@\@citeb \endcsname \relax%
      \ifnum\@tempcnta=\@tempcntb %   Number follows previous--hold on to it
        \ifx\@h@ld\relax%
          \edef \@h@ld{\@citea\csname b@\@citeb\endcsname}%
        \else%
          \edef\@h@ld{\ifmmode{-}\else--\fi\csname b@\@citeb\endcsname}%
        \fi%
      \else%   %  non-successor--dump what's held and do this one
        \@h@ld\@citea\csname b@\@citeb \endcsname%
        \let\@h@ld\relax%
      \fi}%
    \def\@citea{,\penalty\@highpenalty\,}%
  }\@h@ld%
}{#1}}

\makeatother
\begin{document}
\hfuzz=100pt
%\rightmargin -2.75cm
%\textheight 23.0cm
%\topmargin -0.5in
%\baselineskip 16pt
%\parskip 18pt
%\parindent 30pt
%\def\mc{\,\raise -2.truept\hbox{\rlap{\hbox{$\sim$}}\raise5.truept
%\hbox{$<$}\ }}
%\def\Mc{\,\raise -2.truept\hbox{\rlap{\hbox{$\sim$}}\raise5.truept
%\hbox{$>$}\ }}%
%
%
%%%%%%%%%%%%%%%%%%%%%%%%%%%%%%%%%%%%%%%%%%%%%%%%%%%%%%%%%%%%%%%
%
%    List of  the     commands
%%%%%%%%%%%%%%%%%%%%%%%%%%%%%%%%%%%%%%%%%%%%%%%%%%%%%%%%%%%%%%%
%
\newcommand{\be}{\begin{equation}}
\newcommand{\ee}{\end{equation}}
\newcommand{\bea}{\begin{eqnarray}}
\newcommand{\eea}{\end{eqnarray}}
\begin{titlepage}
\makeatletter
\def \thefootnote {\fnsymbol {footnote}} \def \@makefnmark {
\hbox to 0pt{$^{\@thefnmark }$\hss }}
\makeatother
\begin{flushright}
BONN-HE-93-11\\
April, 1993
\end{flushright}
\vspace{2cm}
\begin{center}
{ \large \bf Naked Singularities in Four-dimensional String Backgrounds}\\
\vspace{2cm}
{\large\bf Noureddine Mohammedi} \footnote
{Work supported by the Alexander von Humboldt-Stiftung.}
\footnote{e-mail: nouri@avzw02.physik.uni-bonn.de}
\\
\vspace{.5cm}
\large Physikalisches Institut\\
der Universit\"at Bonn\\
Nussallee 12\\ D-5300 Bonn 1, Germany\\

\baselineskip 18pt
\vspace{.2in}
\vspace{1cm}
{\large\bf Abstract}

\end{center}
It is shown that gauged nonlinear sigma models can be always deformed
by terms proportional to the field strength of the gauge fields
(nonminimal gauging). These deformations can be interpreted as
perturbations, by marginal operators, of conformal coset models.
When applied to the $SL(2,R)\times SU(2)/\left(U(1)\times U(1)\right)$
WZWN model, a large class of four-dimensional curved spacetime
backgrounds are obtained. In particular, a naked singularity may
form at a time when the volume of the universe is different from zero.
%\\
\setcounter {footnote}{0}
\end{titlepage}
\baselineskip 20pt

\section{Introduction}

It is well-established by now that string propagation
on curved spacetimes is described by a two-dimensional
nonlinear sigma model. The conformal invariance
conditions (the equations for the vanishing of the beta
functions) for the sigma model are then interpreted as
defining consistent backgrounds for string
propagation [1,2]. These equations, however, are in general
not easy to solve, especially when they include
higher order terms in perturbation theory. One of
the few techniques for finding these backgrounds is
provided by non-compact gauged WZWN models
(non-compact cosets) [3]. This is mainly because
WZWN models, which are two-dimensional sigma models,
are conformally invariant to all orders in
perturbation theory [4,5].
\par
The programme for constructing curved spacetimes
for string theories received a tremendous boost
when a two-dimensional black hole was discovered
in the $SL(2,R)/U(1)$ coset model [6] and as solution
to the string beta functions [7,8]. Since then many
other papers have used gauged WZWN models to give
one-loop solutions to the conformal invariance
conditions [9-26]. Some of these solutions are related
to each other by a kind of duality transformations [27-34].
An exact form for the backgrounds, consistent
with the loop expansion [35,36], are also found using the
so called ``operator approach'' [10,37,38].
\par
This interest in coset models has subsequently
led to constructing string theories on
four-dimensional curved spacetime backgrounds
and analysing their cosmological implications [39-43].
(Cosmological string backgrounds were also considered
in another context in [44-48]). One of such models is the coset
$SL(2,R)\times SU(2)/\left(U(1)\times U(1)\right)$,
investigated by Nappi
and Witten at the one-loop level [42] and its exact
backgrounds were recently found by Bars and Sfetsos [43].
This model describes a closed, inhomogeneous expanding
and recollapsing universe in four dimensions. The main
feature of this model, however, is the formation of
a naked singularity at the time when the universe
collapses. This last result was also confirmed in [49].
\par
The purpose of this paper is to investigate the effects
of adding, to the gauged
$SL(2,R)\times SU(2)/\left(U(1)\times U(1)\right)$
model, terms proportional to the field strength of the
gauge fields. This alters drastically the singularity
structure of the four-dimensional model. We find, in
particular, that a naked singularity may
form as well when the volume of the universe is different
from zero (the universe is not in a collapsed state).
This might be a possible candidate for the violation of
cosmic censorship.
\par
The reason behind adding these extra terms to the
$SL(2,R)\times SU(2)/\left(U(1)\times U(1)\right)$
coset model, and
to any gauged WZWN model in general, has its roots in the mathematical
formalism of gauging an isometry subgroup of a general
nonlinear sigma model [50,51]. These extra terms are also a neat
and mathematically appealing method for incorporating
marginal perturbations of coset models [52,53].
\par
In section two, we give the formalism for gauging two
abelian isometries of a nonlinear sigma model with
an antisymmetric field. We show that the usual
minimal coupling of the gauge fields (replacing
ordinary derivatives by gauge covariant derivatives)
does not lead to a gauge invariant theory. The gauge
invariant action is instead found by an explicit
use of Noethers's method. At the end of this
section we show how the extra terms, involving
the field strength of the gauge fields, arise in this
formalism. We apply this formalism, in section three, to
the $SL(2,R)\times SU(2)/\left(U(1)\times U(1)\right)$
coset model. We obtain in this
manner a four-dimensional curved spacetime whose singularity
structure is discussed in the last section.

\section{Gauging Two Abelian Isometries of a Sigma Model}

In order to get a four-dimensional manifold from the
$SL(2,R)\times SU(2)$ WZWN model, one has to gauge two abelian
isometries of this group. Here we examine the general formalism
of gauging any two abelian isometries of a general sigma model.
The action for a general bosonic two-dimensional nonlinear sigma
model is given by
\be
S={1\over 4\pi}\int {\rm {d}}^2x\left(\sqrt {\gamma}
\gamma^{\mu\nu}G_{ij}
+\epsilon^{\mu\nu}B_{ij}\right)\partial_\mu\phi^i
\partial_\nu\phi^j\,\,\,.
\ee
The metric $G_{ij}$ and the antisymmetric tensor $B_{ij}$ are the
massless modes of the bosonic string theory. This action is
 invariant under
the infinitesimal global isometries
\be
\delta\phi^i=\varepsilon K^i(\phi)+\widetilde\varepsilon
\widetilde K^i(\phi)
\ee
provided that $K^i$ and $\widetilde K^i$ are Killing vectors of
the metric $G_{ij}$, $\nabla_{(i}K_{j)}=\nabla_{(i}
\widetilde K_{j)}=0$, and the antisymetric tensor $B_{ij}$
satisfies
\bea
\partial_l B_{ij}K^l +B_{lj}\partial_iK^l+B_{il}\partial_jK^l=
\nabla_iL_j -\nabla_jL_i\nonumber\\
\partial_l B_{ij}\widetilde K^l +B_{lj}\partial_i\widetilde K^l
+B_{il}\partial_j\widetilde K^l=
\nabla_i\widetilde L_j -\nabla_j\widetilde L_i
\eea
for some two vectors $L_i$ and $\widetilde L_i$ [54]. These
last two
equations are consequences of the invariance of the action
(but not the Lagrangian) under
\be
B_{ij}\rightarrow B_{ij} +\nabla_{[i}V_{j]}\,\,\,.
\ee
\par
We would like now to gauge the above infinitesimal transformations
(the parameters $\varepsilon$ and $\widetilde\varepsilon$ would
now be local parameters $\varepsilon (x)$
and $\widetilde\varepsilon(x)$). For this gauging to take place,
the generators of the two abelian isometries must commute
\be
\left[K^i\partial_i\,,\,\widetilde K^j\partial_j\right]=0\,\,\,.
\ee
We also introduce two abelian gauge fields $A_{\mu}$ and
$\widetilde A_{\mu}$ transforming as
\be
\delta A_{\mu}= -\partial_\mu\varepsilon\,\,,\,\,
\delta\widetilde A_{\mu}= -\partial_\mu\widetilde\varepsilon\,\,\,.
\ee
Usually, the gauging is carried out by replacing ordinary derivatives
by gauge covariant derivatives (minimal coupling)
\be
\partial_\mu\phi \rightarrow D_\mu\phi^i=\partial_\mu\phi
+A_\mu K^i+\widetilde A_\mu\widetilde K^i\,\,\,.
\ee
However, here and due to the presence of the antisymmetric tensor
$B_{ij}$, the minimal coupling covariantisation will not lead to
a gauge invariant theory [50,51].
\par
We therefore postulate the most general action involving the gauge
 fields
$A_{\mu}$ and $\widetilde A_{\mu}$ to have the form
(Noether's method)
\bea
S_{gauged}&=&{1\over 4\pi}\int d^2x\left\{\sqrt {\gamma}
\gamma^{\mu\nu}G_{ij}
D_\mu\phi^iD_\nu\phi^j+\epsilon^{\mu\nu}B_{ij}
\partial_\mu\phi^i\partial_\nu\phi^j
-2\epsilon^{\mu\nu}C_iA_\mu\partial_\nu\phi^i\right.
\nonumber\\
&-&\left.2\epsilon^{\mu\nu}\widetilde C_i\widetilde A_
\mu\partial_\nu\phi^i
-2\epsilon^{\mu\nu}H(\phi)A_\mu\widetilde A_\nu \right\}
\,\,\,.
\eea
The commutation relation (2.5) insures that the gauge
covariant derivatives
transforms in the correct way
\be
\delta(D_\mu\phi^i)=\left(\varepsilon\partial_jK^i
+\widetilde\varepsilon\partial_j\widetilde K^i\right)
D_\mu\phi^j
\ee
and this guarantees the gauge invariance of the first
term in the action (2.8).
The quantities $C_i(\phi)$, $\widetilde C_i(\phi)$ and
$H(\phi)$ are then determined by requiring gauge invariance of
the rest of the action. First of all,
$C_i(\phi)$ and  $\widetilde C_i(\phi)$ are defined by
\bea
C_i&=&B_{ij}K^j+L_i\nonumber\\
\widetilde C_i&=&B_{ij}\widetilde K^j+\widetilde L_i\,\,\,.
\eea
Secondly, the Lie derivatives of $C_i(\phi)$ and  $\widetilde
C_i(\phi)$
must vanish along the directions of $K^i$ and $\widetilde K^i$
\bea
&\partial_jC_iK^j+C_j\partial_iK^j=0&\nonumber\\
&\partial_j\widetilde C_i\widetilde K^j+\widetilde
 C_j\partial_i
\widetilde K^j=0&\nonumber\\
&\partial_jC_i\widetilde K^j+C_j\partial_i\widetilde K^j=0&
\nonumber\\
&\partial_j\widetilde C_iK^j+\widetilde C_j\partial_iK^j=0&
\eea
Thirdly, we have the following constraints on
$C_i(\phi)$ and  $\widetilde C_i(\phi)$
\bea
&C_jK^j=0&\nonumber\\
&\widetilde C_j\widetilde K^j=0&
\eea
Finally, the function $H(\phi)$ is determined from the equations
\bea
&C_i\widetilde K^i - H=0&\nonumber\\
&\widetilde C_iK^i + H =0&\,\,\,.
\eea
\par
The main feature of the above formalism of gauging a nonlinear sigma
model with a WZ term is the following : By acting with $\partial_i$
on both sides of the two equations in (2.12) and then substituting for
$C_j\partial_iK^j$ and $\widetilde C_j\partial_i\widetilde K^j$ in
the first two equations in (2.11), we find
\bea
&\left(\partial_jC_i-\partial_iC_j\right)K^j=0&\nonumber\\
&\left(\partial_j\widetilde C_i
-\partial_i\widetilde C_j\right)\widetilde K^j=0&\,\,\,.
\eea
Notice that these last two equations remain invariant
under the replacements
\bea
&C_i\rightarrow C_i+\partial_iX&\nonumber\\
&\widetilde C_i\rightarrow \widetilde C_i
+\partial_i\widetilde X&\,\,\,.
\eea
This means that if $C_i$ and $\widetilde C_i$ are solutions to the
above gauge invariance conditions, then $C_i+\partial_iX$ and
$\widetilde C_i+\partial_i\widetilde X$ are solutions, too.
The only restriction on the two functions $X(\phi)$ and
$\widetilde X(\phi)$ come from the requirement that
the rest of the equations involving $C_i$ and
$\widetilde C_i$ should remain invariant under these
replacements. This is so,
provided that
\bea
&\partial_iXK^i=\partial_iX\widetilde K^i=0&
\nonumber\\
&\partial_i\widetilde X\widetilde K^i=
\partial_i\widetilde X K^i=0&\,\,\,.
\eea
The above freedom in determining $C_i$ and $\widetilde C_i$
reflects simply the invariance of the equations (2.3) under
the shifts $L_i\rightarrow L_i+\partial_iX$
and $\widetilde L_i\rightarrow \widetilde L_i
+\partial_i\widetilde X$.
\par
At the level of the action, the above replacement amounts
to adding the following extra terms
\be
S_{extra}={1\over 4\pi}\int d^2x \epsilon^{\mu\nu}\left
(F_{\mu\nu}X(\phi)
+\widetilde F_{\mu\nu}\widetilde X(\phi)\right)\,\,\,,
\ee
where $F_{\mu\nu}$ and $\widetilde F_{\mu\nu}$ are the abelian
curvatures corresponding to the gauge fields $A_\mu$
and $\widetilde A_\mu$, respectively.
The appearance of this additional term is also motivated from
a completely different point of view: In the quantum
theory of the action (2.8), where the gauge fields are
treated as fixed backgrounds, divergences proportional
to $F_{\mu\nu}$ and $\widetilde F_{\mu\nu}$ are unavoidably
generated [50]. Hence for a renormalisable theory we should
add terms proprotional to $F_{\mu\nu}$ and $\widetilde F_{\mu\nu}$
to the classical action in order to absorb these divergences
(see [50] for more details).

\section{The Four-dimensional Curved Spacetime}

The four-dimensional target space is obtained by taking a WZWN
model defined on the group manifold $G=SL(2,R)_{k'}\times
SU(2)_k/H$, where $H$ is a two-dimensional abelian subgroup
of the group of isometries. The central charge for such a model is
\be
c={3k'\over k'+2}+{3k\over k+2}-2
\ee
In the limit of very large $k$ and $k'$, the central charge is,
as required, equal to four (see [42]).
\par
If $(g_1,g_2)$ $\in$ $SL(2,R)_{k'}\times SU(2)_k$, then the subgroup
is chosen to be infinitesimally generated by [42]
\bea
\delta g_1&=&\varepsilon\sigma_3g_1+\left(\widetilde\varepsilon\cos
\alpha+\epsilon\sin\alpha\right)g_1\sigma_3\nonumber\\
\delta g_2&=&i\widetilde\varepsilon\sigma_2g_2
+i\left(-\widetilde\varepsilon\sin
\alpha+\epsilon\cos\alpha\right)g_2\sigma_2\,\,\,.
\eea
The local gauge parameters are $\varepsilon$ and $\widetilde
\varepsilon$, while $\alpha$ is an arbitrary constant and
$\sigma_i$, $i=1,\dots,3$, are the usual $2\times 2$ Pauli
matrices ($\sigma^{\ast}_2=-\sigma_2$). These transformations
generate an anomaly-free subgroup only when $k'=-k$ [42].
\par
We would like now to apply the formalism of the previous
section to the WZWN model defined on the group manifold $G$.
For this purpose, we parametrize the $SL(2,R)$ and the $SU(2)$
groups by
\bea
g_1&=&\left( \begin{array}{cc}
a&u\\
-v&b
\end{array} \right)\,\,\,,\,\,\,ab+uv=1\nonumber\\
g_2&=&\exp\left({i\over 2}(\rho+\lambda)\sigma_2\right)
\exp\left(is\sigma_3\right)
\exp\left({i\over 2}(\rho-\lambda)\sigma_2\right)\,\,\,.
\eea
It is then easy to read off the Killing vectors from the
expressions of
$\delta g_1$ and $\delta g_2$. These are given by
(our six coordinates are $a,u,v,\rho,\lambda,s$)
\bea
K^a&=&(1+\sin\alpha)a\,\,,\,\, K^u=(1-\sin\alpha)u\,\,,\,\,
K^v=-(1-\sin\alpha)v\,\,\,,\nonumber\\
K^\rho&=&\cos\alpha\,\,,\,\,K^\lambda=-\cos\alpha\,\,,\,\,
K^s=0\nonumber\\
\widetilde K^a&=&a\cos\alpha\,\,,\,\,\widetilde K^u=-u\cos\alpha
\,\,,\,\,
\widetilde K^v=v\cos\alpha\,\,\,,\nonumber\\
\widetilde K^\rho&=&(1-\sin\alpha)\,\,,\,\,
\widetilde K^\lambda=(1+\sin\alpha)\,\,,\,\,
\widetilde K^s=0\,\,\,.
\eea
Our aim now is to see whether it is possible to find two
functions $X$ and $\widetilde X$ satisfying the gauge conditions
in (2.16). Using the above Killing vectors and (2.16), we find
\bea
&\left[u\cos\alpha\partial_u-v\cos\alpha\partial_v
-(1+\sin\alpha)\partial_\lambda\right]X=0&\nonumber\\
&\left[a\cos\alpha\partial_a+(1-\sin\alpha)
\partial_\rho\right]X=0&
\eea
and two similar equations for $\widetilde X$. The general solution
to these differential equations is given by
\be
X=X\left(ae^{-{\cos\alpha\over 1-\sin\alpha}\rho}\,,\,
u^\beta v^\gamma e^{{\cos\alpha\over 1+\sin\alpha}(\beta-\gamma)
\lambda}\,,\,
u^{\beta'} v^{\gamma'} e^{{\cos\alpha\over 1+\sin\alpha}
(\beta'-\gamma')
\lambda}\,,\,s\right)\,\,\,,
\ee
where the parameters $(\beta,\beta',\gamma,\gamma')$ satisfy
\be
\beta\gamma'-\beta'\gamma \neq 0\,\,\,.
\ee
A similar expression holds for $\widetilde X$.
As expected $X$ and $\widetilde X$ depend on four variables only
(instead of six). This is because the gauging reduces the
number of degrees of freedom by two.
Notice also that the arguments of $X$ and $\widetilde X$
are invariant under the gauge transformations (3.2).
Hence, $X$ and $\widetilde X$ are functions of the invariants
(under the gauge transformations (3.2)) of the group
$SL(2,R)\times SU(2)$.
\par
Since the two functions $X$ and $\widetilde X$ have been found,
we can therefore proceed to construct our gauged sigma model.
Introducing two abelian gauge fields $A_\mu$ and $\widetilde A_\mu$
and taking $k'=-k$, the nonminimal gauged WZWN action takes the
form
\bea
S_{gauged}+S_{extra}&=&-kS_{WZWN}(g_1)+kS_{WZWN}(g_2)
\nonumber\\
&+&{k\over 2\pi}\int d^2z\left\{A_ztr\left[
\bar\partial g_1g_1^{-1}\sigma_3\right]+
\left(\widetilde A_{\bar z}\cos\alpha+A_{\bar z}\sin\alpha
\right)tr\left[g_1^{-1}\partial g_1\sigma_3\right]\right\}
\nonumber\\
&-&{ik\over 2\pi}\int d^2z\left\{\widetilde A_ztr\left[
\sigma_2\bar\partial g_2g_2^{-1}\right]+
\left( A_{\bar z}\cos\alpha-\widetilde A_{\bar z}\sin\alpha
\right)tr\left[\sigma_2g_2^{-1}\partial g_2\right]\right\}
\nonumber\\
&+&{k\over 2\pi}\int d^2z\left(A_z\widetilde A_{\bar z}
\cos\alpha + A_zA_{\bar z}\sin\alpha\right)tr
\left[\sigma_3g_1\sigma_3g_1^{-1}\right]
\nonumber\\
&+&{k\over 2\pi}\int d^2z\left(\widetilde A_z A_{\bar z}
\cos\alpha -\widetilde A_z\widetilde A_{\bar z}
\sin\alpha\right)tr
\left[\sigma_2g_2\sigma_2g_2^{-1}\right]
+{k\over \pi}\int d^2z\left(A_zA_{\bar z}+
\widetilde A_{\bar z}\widetilde A_z\right)
\nonumber\\
&+&{k\over 2\pi}\int d^2z\left\{\left(\bar\partial A_z
 - \partial
A_{\bar z}\right)X+\left(\bar\partial\widetilde A_z
- \partial
\widetilde A_{\bar z}\right)\widetilde X\right\}\,\,\,.
\eea
The standard WZWN action is
\be
S_{WZWN}(g)=-{1\over 4\pi}\int_\Sigma d^2z tr
\left[(g^{-1}\partial g)(g^{-1}\bar\partial g)\right]
-{i\over 12\pi}\int_B tr\left[(g^{-1}dg)\wedge
(g^{-1}dg)\wedge(g^{-1}dg)\right]\,\,\,.
\ee
Using our parametrization for $g_1$ and $g_2$, the gauged
WZWN action leads to a gauged nonlinear sigma model of the
form (2.8) and it is only for $k'=-k$ that all the
gauge invariance conditions of the previous section are
fullfiled.
\par
We would like now to fix a gauge and integrate out the gauge
fields.
A suitable gauge choice is found by taking $g_1$ to be given
by
\be
g_1=\left(\begin{array}{cc}
\cos\psi &\sin\psi \\
-\sin\psi &\cos\psi
\end{array}\right)\,\,\,.
\ee
The integration over the gauge fields leads to a nonlinear sigma
model having the following space-time metric
\bea
DG_{\psi \psi }&=&2\left(1-\cos 2\psi \cos 2s\right)
+2\sin\alpha\left(\cos 2\psi -\cos2s\right)
+\left(\sin\alpha\cos 2s -1\right)(\partial_\psi X)^2
\nonumber\\
&-&\left(\sin\alpha\cos 2\psi  +1\right)(\partial_\psi
\widetilde X)^2
+\cos\alpha\left(\cos 2s+\cos 2\psi \right)\partial_\psi X
\partial_\psi  \widetilde X
\nonumber\\
DG_{\psi  s}&=&
\left(\sin\alpha\cos 2s -1\right)\partial_\psi
X\partial_sX
-\left(\sin\alpha\cos 2\psi  +1\right)\partial_\psi
\widetilde X\partial_s\widetilde X\nonumber\\
&+&{1\over 2}\cos\alpha\left(\cos 2s+\cos 2\psi \right)
\left(\partial_\psi X\partial_s\widetilde X+
\partial_sX\partial_\psi  \widetilde X\right)
\nonumber\\
DG_{\psi \rho}&=&
\left(\sin\alpha\cos 2s -1\right)\partial_\psi X
\partial_\rho X
-\left(\sin\alpha\cos 2\psi  +1\right)\partial_\psi
\widetilde X\partial_\rho\widetilde X
\nonumber\\
&+&{1\over 2}\cos\alpha\left(\cos 2s+\cos 2\psi \right)
\left(\partial_\psi X\partial_\rho\widetilde X+
\partial_\rho X\partial_\psi  \widetilde X\right)
\nonumber\\
&+&\cos^2s\left(\cos 2\psi +1\right)\left[-cos\alpha
\partial_\psi  X
+\left(\sin\alpha +1\right)\partial_\psi \widetilde
 X\right]
\nonumber\\
DG_{\psi \lambda}&=&
\left(\sin\alpha\cos 2s -1\right)\partial_\psi X\partial_
\lambda X
-\left(\sin\alpha\cos 2\psi  +1\right)\partial_\psi
\widetilde X\partial_\lambda\widetilde X
\nonumber\\
&+&{1\over 2}\cos\alpha\left(\cos 2s+\cos 2\psi \right)
\left(\partial_\psi  X\partial_\lambda\widetilde X+
\partial_\lambda X\partial_\psi  \widetilde X\right)
\nonumber\\
&+&\sin^2s\left(\cos 2\psi -1\right)\left[-\cos\alpha
\partial_\psi  X
+\left(\sin\alpha-1\right)\partial_\psi \widetilde
 X\right]
\nonumber\\
DG_{ss}&=&-2\left(1-\cos 2\psi \cos 2s\right)
-2\sin\alpha\left(\cos 2\psi -\cos2s\right)
+\left(\sin\alpha\cos 2s -1\right)(\partial_s X)^2
\nonumber\\
&-&\left(\sin\alpha\cos 2\psi  +1\right)(\partial_s
\widetilde X)^2
+\cos\alpha\left(\cos 2s+\cos 2\psi \right)\partial_s X
\partial_s \widetilde X
\nonumber\\
DG_{s\rho}&=&
\left(\sin\alpha\cos 2s -1\right)\partial_s X
\partial_\rho X
-\left(\sin\alpha\cos 2\psi  +1\right)\partial_s
\widetilde X\partial_\rho\widetilde X
\nonumber\\
&+&{1\over 2}\cos\alpha\left(\cos 2s+\cos 2\psi
 \right)
\left(\partial_s X\partial_\rho\widetilde X+
\partial_\rho X\partial_s  \widetilde X\right)
\nonumber\\
&+&\cos^2s\left(\cos 2\psi +1\right)\left[-\cos
\alpha\partial_s  X
+\left(\sin\alpha+1\right)\partial_s \widetilde
 X\right]
\nonumber\\
DG_{s\lambda}&=&
\left(\sin\alpha\cos 2s -1\right)\partial_s X
\partial_\lambda X
-\left(\sin\alpha\cos 2\psi  +1\right)\partial_s
\widetilde X\partial_\lambda\widetilde X
\nonumber\\
&+&{1\over 2}\cos\alpha\left(\cos 2s+\cos 2\psi \right)
\left(\partial_s X\partial_\lambda\widetilde X+
\partial_\lambda X\partial_s  \widetilde X\right)
\nonumber\\
&+&\sin^2s\left(\cos 2\psi -1\right)\left[-\cos\alpha
\partial_s  X
+\left(1-\sin\alpha\right)\partial_s \widetilde X\right]
\nonumber\\
DG_{\rho\rho}&=&
-4\cos^2s\cos^2\psi(1+\sin\alpha) +\left(\sin\alpha\cos 2s
-1\right)
\left(\partial_\rho X\right)^2
-\left(\sin\alpha\cos 2\psi +1\right)
\left(\partial_\rho \widetilde X\right)^2\nonumber\\
&+&\cos\alpha\left(\cos 2s +\cos 2\psi\right)
\partial_\rho X\partial_\rho \widetilde X
\nonumber\\
&+&2\cos^2s\left(\cos 2\psi +1\right)\left[-\cos
\alpha\partial_\rho X
+\left(\sin\alpha +1\right)
\partial_\rho \widetilde X\right]
\nonumber\\
DG_{\rho\lambda}&=&
\left(\sin\alpha\cos 2s -1\right)\partial_\rho X
\partial_\lambda X
-\left(\sin\alpha\cos 2\psi +1\right)\partial_\rho
\widetilde X\partial_\lambda \widetilde X
\nonumber\\
&+&{1\over 2}\cos\alpha\left(\cos 2s +
\cos 2\psi\right)
\left(\partial_\rho X\partial_\lambda\widetilde X+
\partial_\lambda X\partial_\rho\widetilde X\right)
\nonumber\\
&+&\cos^2s\left(\cos 2\psi +1\right)\left[
-\cos\alpha\partial_\lambda X
+\left(\sin\alpha +1\right)
\partial_\lambda \widetilde X\right]
\nonumber\\
&+&\sin^2s\left(\cos 2\psi -1\right)\left[-\cos\alpha
\partial_\rho X
+\left(\sin\alpha -1\right)
\partial_\rho \widetilde X\right]
\nonumber\\
DG_{\lambda\lambda}&=&
-4\sin^2s\sin^2\psi(1-\sin\alpha) +\left(\sin\alpha\cos 2s
 -1\right)
\left(\partial_\lambda X\right)^2
-\left(\sin\alpha\cos 2\psi +1\right)
\left(\partial_\lambda \widetilde X\right)^2
\nonumber\\
&+&\cos\alpha\left(\cos 2s +\cos 2\psi\right)
\partial_\lambda X\partial_\lambda\widetilde X
\nonumber\\
&+&2\sin^2s\left(\cos 2\psi -1\right)\left[-\cos\alpha
\partial_\lambda X
+\left(\sin\alpha-1\right)
\partial_\lambda \widetilde X\right]\,\,\,.
\eea
The factor $D$ determines the singularities of the metric and
is given by
\be
D=-{4\pi\over k}\left[\left(1-\cos 2\psi \cos 2s\right)
+\sin\alpha\left(\cos 2\psi -\cos2s\right)\right]\,\,\,.
\ee
Notice that if $\psi =0$ and $s=0$ or $\psi={\pi\over 2}$ and
$s={\pi\over 2}$, then $D$ vanishes and the metric is singular.
These singularities are interpreted, in the next section, as
naked singularities.

\section{Discussions}

Let us now discuss the cosmological model in which the time parameter
is $\psi$, taking values in the range $0\le\psi\le{\pi\over 2}$. We
distinguish therefore two different cases :
\par
The first is when $X$ and $\widetilde X$ are set to constants.
This case was elaborated in details in ref.[42]. Here we give a
brief summary. The target space line element in this case
is given by
\be
dl^2=-d\psi^2+ds^2-{4\over D}\left
((1+\sin\alpha)\cos^2s\cos^2\psi d\rho^2
+(1-\sin\alpha)\sin^2s\sin^2\psi d\lambda^2\right)\,\,\,.
\ee
The universe starts from a collapsed state
(a big bang) at $\psi=0$, since the determinant of the metric
$G_{ij}$ vanishes (hence the volume of the universe). At $\psi
={\pi\over 2}$, the universe collapses again (a big crunch).
More appealing is
the presence of a possible candidate for a naked singularity
and hence a possible violation of the cosmic censorship. Indeed, at
$\psi=s=0$ or $\psi=s={\pi\over 2}$ the metric is not defined
($D=0$) and
since these singularities appear at definite values of the time
$\psi$, they might stand as candidates for naked singularities.
However, this is not the case since it is precisely at
$\psi=0$ or $\psi={\pi\over 2}$ that the universe is in a collapsed
state. Therefore the naked singularities cannot be seen as the whole
universe collapses when they would have appeared.
\par
The second case is when $X$ and $\widetilde X$ are not constants
(or at least one of the two is not). We would like to examine
the fate of the above singularities when the action
contains the extra nonminimal term.
There is no reason, a priori, for the volume of the universe
to vanish at $\psi =0$ since the functions $X$ and
$\widetilde X$ can be chosen at will. We will provide, in
what follows, an example in which the universe is not in a
collapsed state at $\psi =0$.
\par
Notice that all the extra terms appearing in the metric (3.11)
involve only the derivatives of $X$ and $\widetilde X$. Let us
therefore write down these derivatives in the gauge specified
by (3.10). We have for the derivatives of $X$ (and similarly
for $\widetilde X$)
\bea
\partial_\psi X&=& -e^{-{\cos\alpha\over 1-\sin\alpha}
\rho}(\sin\psi)
\partial_{\xi_1}X
+(\beta +\gamma)e^{{\cos\alpha\over 1+\sin\alpha}(\beta
-\gamma)\lambda}(\cos\psi)\left(\sin\psi\right)^
{\beta+\gamma -1}
\partial_{\xi_2}X \nonumber\\
&+&
(\beta' +\gamma')e^{{\cos\alpha\over 1+\sin\alpha}
(\beta' -\gamma')\lambda}(\cos\psi)\left(\sin\psi\right)
^{\beta'+\gamma' -1}
\partial_{\xi_3}X \nonumber\\
\partial_\rho X&=& -{\cos\alpha\over 1-\sin\alpha}
e^{-{\cos\alpha\over 1-\sin\alpha}\rho}(\cos\psi)
\partial_{\xi_1}X \nonumber\\
\partial_\lambda X&=&
{\cos\alpha\over 1+\sin\alpha}\left[(\beta
-\gamma)
e^{{\cos\alpha\over 1+\sin\alpha}(\beta
-\gamma)\lambda}
\left(\sin\psi\right)^{\beta+\gamma }
\partial_{\xi_2}X \right.\nonumber\\
&+&\left.
(\beta' -\gamma')
e^{{\cos\alpha\over 1+\sin\alpha}(\beta'
-\gamma')\lambda}
\left(\sin\psi\right)^{\beta'+\gamma'}
\partial_{\xi_3}X\right] \nonumber\\
\partial_s X&=&\partial_{\xi_4} X\,\,\,.
\eea
Here, the coordinates  $\xi_1$,  $\xi_2$,
$\xi_3$ and $\xi_4$ are defined by
\bea
\xi_1&=&e^{-{\sin\alpha\over 1-\sin\alpha}\rho}
\cos\psi \,\,,\,\,
\xi_2=e^{{\cos\alpha\over 1+\sin\alpha}(\beta'
-\gamma')\lambda}
\left(\sin\psi\right)^{\beta+\gamma}
\,\,,
\nonumber\\
\xi_3&=&e^{{\cos\alpha\over 1+\sin\alpha}(\beta
 -\gamma)\lambda}\left(\sin\psi\right)^{\beta'
+\gamma'}\,\,,\,\,
\xi_4=s\,\,\,.
\eea
\par
The volume of the universe is different from zero at $\psi=0$
only if all the derivatives of  $X$ and $\widetilde X$
evaluated at $\psi=0$ are well-defined
and different from zero. This situation can always be realized
by suitably choosing $X$ and $\widetilde X$. For example, by taking
\be
\beta +\gamma =1\,\,,\,\,\beta' +\gamma' =0\,\,\,,
\ee
we see that all the derivatives of $X$ and $\widetilde X$ are
different from zero at $\psi=0$ provided that $\partial_{\xi_i}X$ and
$\partial_{\xi_i}\widetilde X$, $i=1,\dots,4$, are
well-defined and different from zero at $\psi=0$.
\par
Therefore, we can always choose $\partial_{\xi_i}X$ and
$\partial_{\xi_i}\widetilde X$ in such a way that the volume
of the universe is different from zero at $\psi=0$.
This statement holds for generic values of $s$. It is also easy
to explicitly verify that if none of the derivatives of
$X$ and $\widetilde X$ vanishes at $\psi =0$ and $s=0$, then all
the components of the metric in (3.11) are nonvanishing at
$\psi =0$ and $s=0$. Hence, we have provided an example in which
a kind of naked singularity might form when the universe is not in a
collapsed state. This is in contrast to the results reached in
[42,43,49], where the universe was found to collapse right at
the time when a naked singularity was about to form. We were
not able to prove that the universe does not collapse at
$\psi={\pi\over 2}$ and $s={\pi\over 2}$.
\par
We would like now to address some issues related to our model.
The first issue regards the gauge fixing. The gauge choice (3.10)
covers only a patch of space-time and it is possible to
continue past
$\psi=0$ and $\psi={\pi\over 2}$. The continuation past $\psi=0$ is
made by taking the gauge condition [42]
\be
g_1={1\over 1+x}\left(\begin{array}{cc}
1&1\\
-x&1
\end{array}\right)\,\,\,.
\ee
The resulting target space-time metric is yet another complicated
expression of the form (3.11). What is of interest to us here is
the equivalent of the denominator $D$. This is given by
\be
D=-{8\pi\over k}{1\over 1+x}\left[
(1-x)\cos 2s -(1+x) +\sin\alpha\left((1+x)\cos 2s
-(1-x)\right)
\right]\,\,\,.
\ee
Hence the metric is not defined for $x=0$ and $s=0$. Here also it
is possible to find $X$ and $\widetilde X$ such that the volume
of the universe is different from zero at $x=0$ ans $s=0$. We
mention here that none of the two chosen gauges covers the
whole four-dimensional space-time (see ref.[43]). Therefore,
the issue of gauge fixing in this model, and in gauged
WZWN models in general, needs further examinations.
\par
The second problem concerns the dilaton field in this model.
So far, we have given only the metric field (the antisymmetric
tensor can be extracted in the same way as the metric). Solving
the one-loop beta functions using the metric (3.11) is not an easy
task. However, this task could be made somehow simpler by putting
some restrictions on the functions $X$ and $\widetilde X$. Since
$X$ and $\widetilde X$ are scalars, we could require them
to satisfy the equations of a massless scalar on the target
space,
\be
\nabla^2X=\nabla^2\widetilde X=0\,\,\,.
\ee
The Laplacian is evaluated using the metric (4.1). This requirement
is equivalent to demanding the vanishing of the one-loop
counterterms proportional to $F_{\mu\nu}$ and $\widetilde F_{\mu\nu}$
which would arise from the action (2.8) when the gauge fields are
treated  as backgrounds [50,51].
\par
In the language of conformal field theory, and using the techniques of
ref.[10], this is the same as requiring $X$ and $\widetilde X$
to be conformal operators of dimension $(0,0)$ so that the
extra terms in the action (3.8) are of dimension $(1,1)$. This
provides a way of analysing perturbations, by marginal operators,
in coset theories. This is an interesting issue to which
we hope to return.
\par
Finally, it is worth mentioning that the singularities of the
$SL(2,R)\times SU(2)/\left(U(1)\times U(1)\right)$ model
(without the extra terms) were boosted away [55] throught the
introduction of a fifth coordinate and by using an $O(3,3)$
transformation. It is therefore of great interest to explore the
effects of duality transformations on our metric.

\vspace{0.5cm}

\paragraph{Acknowledgements:} I would like to thank H. Arfaei, L. Feh\'er,
J. M. Figueroa-O'Farrill and W. Nahm for many useful discusions
and comments. The financial support  from the Alexander
von Humboldt-Stiftung is also hereby acknowledged.

\end{document}